\documentclass[sigconf]{acmart}
\AtBeginDocument{%
  }

\setcopyright{acmlicensed}
\copyrightyear{2018}
\acmYear{2018}
\acmDOI{XXXXXXX.XXXXXXX}
\acmConference[Conference acronym 'XX]{Make sure to enter the correct
  conference title from your rights confirmation email}{June 03--05,
  2018}{Woodstock, NY}
\acmISBN{978-1-4503-XXXX-X/2018/06}

\usepackage{amsmath,amssymb,amsfonts}
\usepackage{algorithmic}
\usepackage{graphicx}
\usepackage{textcomp}
\usepackage{caption}
\usepackage{adjustbox}
\usepackage{prettyref}
\usepackage{multirow}
\usepackage{tabularx}
\usepackage{makecell} 
\usepackage{url}
\usepackage{subcaption}

\newrefformat{sec}{Section~\ref{#1}}
\newrefformat{fig}{Fig.~\ref{#1}}
\newrefformat{tab}{Table~\ref{#1}}
\begin{document}

\title{Mind Your Vision: Multimodal Estimation of Refractive Disorders Using Electrooculography and Eye Tracking}

\author{Xin Wei}
\email{wei.xin.wy0@is.naist.jp}
\orcid{0000-0001-8461-0428}
\affiliation{%
  \institution{Nara Institute of Science and Technology}
  \city{Ikoma}
  \state{Nara}
  \country{Japan}
}

\author{Huakun Liu}
\affiliation{%
  \institution{Nara Institute of Science and Technology}
  \city{Ikoma}
  \state{Nara}
  \country{Japan}
}
\email{liu.huakun.li0@is.naist.jp}

\author{Yutaro Hirao}
\affiliation{%
  \institution{Nara Institute of Science and Technology}
  \city{Ikoma}
  \state{Nara}
  \country{Japan}
}
\email{yutaro.hirao@is.naist.jp}

\author{Monica Perusquia-Hernandez}
\affiliation{%
  \institution{Nara Institute of Science and Technology}
  \city{Ikoma}
  \state{Nara}
  \country{Japan}
}
\email{m.perusquia@is.naist.jp}

\author{Katsutoshi Masai}
\affiliation{%
  \institution{Kyushu University}
  \city{Fukuoka}
  \state{Fukuoka}
  \country{Japan}
}
\email{masai@ait.kyushu-u.ac.jp}

\author{Hideaki Uchiyama}
\affiliation{%
  \institution{Nara Institute of Science and Technology}
  \city{Ikoma}
  \state{Nara}
  \country{Japan}
}
\email{hideaki.uchiyama@is.naist.jp}

\author{Kiyoshi Kiyokawa}
\affiliation{%
  \institution{Nara Institute of Science and Technology}
  \city{Ikoma}
  \state{Nara}
  \country{Japan}
}
\email{kiyo@is.naist.jp}

\renewcommand{\shortauthors}{Wei et al.}

\begin{abstract}
Refractive errors are among the most common visual impairments globally, yet their diagnosis often relies on active user participation and clinical oversight. This study explores a passive method for estimating refractive power using two eye movement recording techniques: electrooculography (EOG) and video-based eye tracking. Using a publicly available dataset recorded under varying diopter conditions, we trained Long Short-Term Memory (LSTM) models to classify refractive power from unimodal (EOG or eye tracking) and multimodal configuration. We assess performance in both subject-dependent and subject-independent settings to evaluate model personalization and generalizability across individuals. Results show that the multimodal model consistently outperforms unimodal models, achieving the highest average accuracy in both settings: 96.207\% in the subject-dependent scenario and 8.882\% in the subject-independent scenario. However, generalization remains limited, with classification accuracy only marginally above chance in the subject-independent evaluations. Statistical comparisons in the subject-dependent setting confirmed that the multimodal model significantly outperformed the EOG and eye-tracking models. However, no statistically significant differences were found in the subject-independent setting. Our findings demonstrate both the potential and current limitations of eye movement data-based refractive error estimation, contributing to the development of continuous, non-invasive screening methods using EOG signals and eye-tracking data.
\end{abstract}



\begin{teaserfigure}
    \centering
  \includegraphics[width=.9\textwidth]{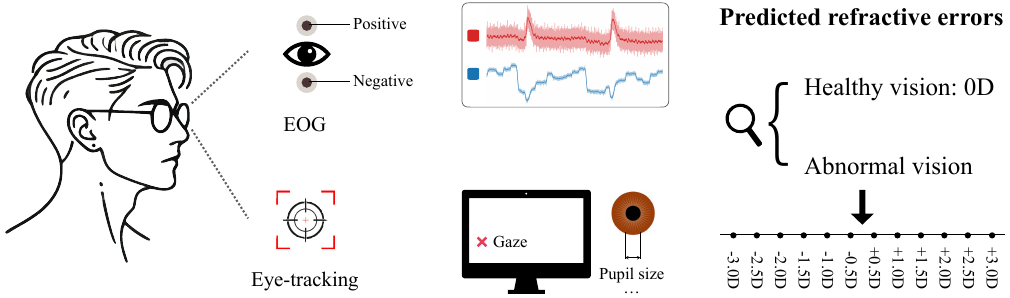}
  \caption{Multimodal classification of refractive error using eye movement data.}
  \Description{Enjoying the baseball game from the third-base
  seats. Ichiro Suzuki preparing to bat.}
  \label{fig:teaser}
\end{teaserfigure}


\maketitle

\section{Introduction}
Refractive errors are one of the leading causes of visual impairment affecting all age groups~\cite{naidoo_global_2016}.
It occurs when the shape of the eye, cornea, or lens prevents light from focusing accurately on the retina, thus transforming vision into a blur~\cite{noauthor_types_nodate}.
The symptoms can impact patients in many aspects, ranging from hindering young children's social, cognitive, and educational development to increasing the risk of falls and reducing independence in older adults~\cite{lanca_editorial_2023, pirindhavellie_impact_2023, ouyang_risk_2022}.
Regular eye examinations are recommended once every one to two years depending on age and risk factors, which may be inaccessible in underdeveloped or rural areas~\cite{noauthor_comprehensive_nodate, rao_patient_2024}.
This limitation, combined with the hard-to-notice nature of early-stage refractive errors, can make them harder to detect.
In recent years, researchers have been working to transform clinic-based visual tests into home-based applications.
A major direction of innovation involves adapting these tests for use on smartphones, computers, and AR/VR devices~\cite{toner_handy_2014, bastawrous_development_2015, hazari_validation_2022, wisse_validation_2019, shekar_vreye_2020}.
However, these assessments remain discrete and heavily depend on user initiative, making them less effective before symptoms or perceived risk arise.
To overcome the need for user engagement, continuous monitoring provides a passive way to identify these disorders before patients are aware of them.
In addition, monitoring the trends in visual function over time can enrich personal health data and support early intervention.

Variations in the cornea, lens, or the shape of the eyeball cause refractive errors.
Several studies have further suggested that these errors can affect not only the anatomical structures of the eyes, but also accommodative behaviors and surrounding facial musculature. 
For example, it has been observed that myopia development involves not only axial elongation but also significant anatomical changes in both the anterior and posterior segments of the eye~\cite{richdale_effect_2016, muralidharan_morphological_2019}.
Abraham et al. reported significant differences in accommodation amplitude, primarily controlled by changes in lens shape and ciliary muscle activity, among emmetropes, hyperopes, and myopes~\cite{abraham_amplitude_2005}.
Zielinski et al. found that individuals with myopia exhibited structural and functional differences in both extraocular and masticatory muscles compared to those without refractive errors~\cite{zielinski_masticatory_2022, zielinski_high_2023}.
These findings indicate that eye movement patterns may differ between individuals with and without refractive errors, and may also change following the onset or progression of such conditions.
Building on this, analyzing eye movements presents a feasible strategy for estimating refractive error.

In this study, we investigate the effectiveness of two eye movement recording techniques, video-based eye tracking and electrooculography (EOG), to support the screening of refractive conditions.
EOG is a low-cost technique that measures the electrical potential between the cornea and retina~\cite{mowrer_corneo-retinal_1935}.
It can infer eye movements from voltage changes in the signal~\cite{marg_development_1951}.
Video-based eye-tracking can capture both pupil and gaze data and rich contextual information, making it effective for analyzing the wearer’s visual behavior~\cite{ebisawa_improved_1998, hua_video-based_2006}.
Moreover, these two techniques can operate without requiring active user input.
This passive nature allows seamless integration into daily life without disrupting the user's routine, thereby greatly enhancing usability.

Using EOG signals for refractive power assessment has been previously explored, but the potential of eye-tracking data for continuous and passive evaluation has not been fully investigated~\cite{wei_unobtrusive_2023, pirso_measuring_2024}.
To fill this gap, We employ unidirectional multi-layer Long Short-Term Memory (LSTM)–based models using both EOG and eye-tracking data to classify refractive power in unimodal and multimodal configurations.
In addition, we consider two application scenarios: within-subject and across-subject.
The within-subject scenario reflects a personalized approach where data collected from the target user is used to train the model for subsequent estimation.
In contrast, the across-subject scenario examines the feasibility of building a generalizable model using data from other users and applying it to a new individual.

The main contributions of this study are:
\begin{itemize}
    \item We propose an LSTM-based framework for refractive power classification using eye movement data.
    \item We compare the performance of using EOG and eye-tracking data for refractive condition assessment in both unimodal and multimodal settings, examining the individual and combined effectiveness of each modality.
    \item We assess the proposed approach in both within-subject and across-subject scenarios, examining its potential for personalized and generalized refractive power estimation.
\end{itemize}

\section{Related Works}

Eye movement is an important criterion in diagnosing visual problems in clinical settings.
A common observation-based approach involves asking individuals to follow moving targets while the clinician visually observes for abnormalities, misalignment, or lack of coordination in eye movements.
However, recent advancements in mobile and sensing technologies have made it possible to measure eye movements anytime and anywhere. 
Furthermore, progress in machine learning has enabled the detection of a broader range of disorders from eye movement data, not limited to oculomotor disorders.

The two eye movement recording techniques used in this study, video-based eye tracking and EOG, have been extensively studied in two main directions.
On one hand, eye movement recognition is widely applied in the field of human-computer interaction (HCI), including applications such as user interface control and adaptive visual displays based on gaze tracking~\cite{jacob_commentary_2003, cantoni_eye_2014, chahine_eye_2023, estrany_human_2008, bulling_wearable_2009}.
It also plays an essential role in assistive technologies that support people with impairments~\cite{ubeda_wireless_2011, narayan_eog-assist_2024, wang_characterizing_2025}.
On the other hand, researchers are increasingly exploring its potential in healthcare.
With growing evidence that eye movement patterns reflect cognitive and mental conditions, researchers are increasingly utilizing metrics such as fixation duration, pupil size, and blink rate to assess mental workload, fatigue, and overall well-being~\cite{tag_continuous_2019, walter_cognitive_2021, ono_scoping_2023}.
In particular, EOG is useful for sleep research because it can record eye movements when the eyes are closed~\cite{fan_eognet_2021}.

Eye movement analysis has also been widely applied in the detection of visual disorders.
Many studies focus on conditions related to oculomotor function.
Chen et al. presented a strabismus detection approach by constructing a gaze deviation image that reflects the misalignment between each eye’s gaze and the target point, which is then processed by a convolutional neural network and classified using a support vector machine~\cite{chen_strabismus_2018}.
Nixon et al. developed and validated STARE, an automated strabismus screening tool that combines augmented reality and eye tracking, which demonstrated 100\% sensitivity and specificity in detecting horizontal strabismus~\cite{nixon_feasibility_2023}.
Also, Varier et al. developed a real-time EOG-based system to detect strabismus and nystagmus, showing that EOG signal patterns can effectively distinguish normal from abnormal eye movements~\cite{varier_electrooculogram_2014}.
In addition to oculomotor dysfunction, Grootjen et al. utilized eye-tracking and machine learning to estimate the likelihood of central field loss based one eye movement data~\cite{grootjen_assessing_2023}.

Focusing on refractive error assessment, Pirso et al. used continuous psychophysics and eye tracking to estimate refractive error through contrast sensitivity during a visual tracking task~\cite{pirso_measuring_2024}. While it shows promise for clinically challenging populations, the approach requires active user engagement and yields only a discrete, momentary assessment. 
Kaya et al. proposed a method for classifying refractive errors using EOG signals recorded from myopic, hyperopic, and emmetropic participants during dark and light adaptation phases. They evaluated several data mining algorithms, including logistic regression, Naive Bayes, random forest, and REP tree, to distinguish among the three group, with the REP tree achieving the highest accuracy of 95.45\%~\cite{kaya_classification_2018}.
Wei et al. simulated different degrees of refractive error using trial lenses and trained a ResNet-18 model to classify these conditions based on EOG recordings obtained during the tasks~\cite{wei_unobtrusive_2023}.
Their subject-dependent models achieved a high average accuracy of 94.7\%, while subject-independent models performed poorly, with a mean accuracy of only 34.5\%.

Building on these efforts, the potential of using eye-tracking data for passive refractive error estimation remains unexplored.
Meanwhile, EOG-based methods have shown promise but face difficulties in generalization across individuals.
Given that EOG is highly sensitive to subtle eye movements, it can capture small variations induced by changes in refractive power~\cite{eggert_eye_2007}.
In contrast, precalibrated optical eye trackers provide more standardized measurements and are generally less affected by inter-individual differences, though factors like eye shape or eyelid occlusion may still introduce some variability~\cite{hansen_eye_2010, linden_calibration_2021}.
By combining these two modalities, it is possible to capture both fine-grained eye movement signals and relatively more robust, standardized gaze features.
In this study, we aim to (1) investigate the feasibility of classifying refractive errors using eye-tracking signals, and (2) explore a multimodal strategy that leverages the complementary strengths of both eye-tracking and EOG to improve performance, particularly in subject-independent settings.

\begin{figure*}[t]
    \centering
    \captionsetup{name={Figure}, justification=justified}
    \includegraphics[width=.9\textwidth]{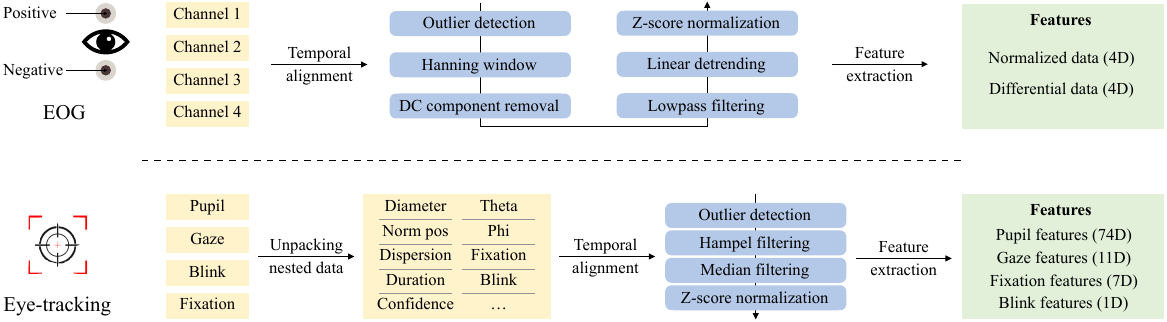}
    \caption{Illustration of the processing techniques. This figure presents a flowchart of the signal processing steps.}
    \label{fig:flowchart}
\end{figure*}

\begin{table*}[h]
\centering
\caption{Summary of Eye-Tracking Features Grouped by Category}
\begin{tabularx}{\textwidth}{|p{3.2cm}|X|}
\hline
\textbf{Feature Group} & \textbf{Features} \\
\hline
\textbf{Pupil Features (Eye 0, 37D)} & 
{\small2D pupil diameter, 2D normalized pupil position (X, Y), 2D ellipse center (X, Y), 2D ellipse axes (A, B), 2D ellipse angle, 2D pupil diameter measured by the 3D detector, 3D pupil diameter, 3D normalized pupil position (X, Y), 3D ellipse center (X, Y), 3D ellipse axes (A, B), 3D ellipse angle, Eyeball sphere center (X, Y, Z), Eyeball sphere radius, 2D projected sphere center (X, Y), 2D projected sphere axes (A, B), 2D projected sphere angle, 3D pupil circle center (X, Y, Z), 3D pupil circle normal (X, Y, Z), 3D pupil circle radius, Circle normal spherical angles (theta, phi), Pupil location (X, Y)} \\
\hline
\textbf{Pupil Features (Eye 1, 37D)} & 
{\small2D pupil diameter, 2D normalized pupil position (X, Y), 2D ellipse center (X, Y), 2D ellipse axes (A, B), 2D ellipse angle, 2D pupil diameter measured by the 3D detector, 3D pupil diameter, 3D normalized pupil position (X, Y), 3D ellipse center (X, Y), 3D ellipse axes (A, B), 3D ellipse angle, Eyeball sphere center (X, Y, Z), Eyeball sphere radius, 2D projected sphere center (X, Y), 2D projected sphere axes (A, B), 2D projected sphere angle, 3D pupil circle center (X, Y, Z), 3D pupil circle normal (X, Y, Z), 3D pupil circle radius, Circle normal spherical angles (theta, phi), Pupil location (X, Y)} \\
\hline
\textbf{Gaze Features (11D)} & 
{\small3D gaze origin (X, Y, Z), 3D gaze direction (X, Y, Z), 3D gaze point (X, Y, Z), 2D normalized gaze position (X, Y)} \\
\hline
\textbf{Fixation Features (7D)} & 
{\small Fixation dispersion, Fixation duration, 2D normalized fixation position (X, Y), 3D fixation point (X, Y, Z)} \\
\hline
\textbf{Blink Features (1D)} & 
{\small Blink type} \\
\hline
\end{tabularx}
\label{tab:eye_tracking_features}
\end{table*}

\section{Methods}
\subsection{Dataset}
We use a dataset comprising EOG and eye-tracking data collected from 37 participants (13 females, mean age = 28.1 years, SD = 4.44) in an experiment designed to support refractive power estimation~\cite{wei_electro_2025}.
Participants’ eye refractive states were manipulated using optometric trial lenses ranging from -3.0 to +3.0 diopters in 0.5-diopter increments, resulting in 13 distinct conditions.
Under each condition, participants completed a sequence of visual tasks, including eight alternating fixation and pursuit trials, followed by one reading task.
Each fixation trial lasted 5 seconds, during which participants stared at a stationary target displayed at the center.
The target positions were selected from a predefined set of eight different locations on the screen, without repetition.
Pursuit trials required participants to follow a moving target along one of four straight-line trajectories—horizontal, vertical, and two diagonal directions—with each trajectory appearing twice per condition.
Trial durations ranged from 1.43 to 2.92 seconds, depending on the trajectory.
Specifically, the target positions for fixation trials and the movement trajectories for pursuit trials were consistent across conditions but presented in a randomized order.
The reading task involved reading an English text for 40 seconds.

EOG signals were recorded using two Shimmer3 ExG units\footnote[1]{https://shimmersensing.com/}.
One Shimmer PROTO3 unit for receiving event triggers and two ExG units were synchronized prior to the experiment.
This setup, using ten electrodes, captured four channels of signals corresponding to the horizontal and vertical eye movements of both eyes at a sampling rate of 512~Hz.
Eye-tracking data were collected using two Pupil Core eye cameras and one scene camera at a sampling rate of 120~Hz~\cite{kassner_pupil_2014}.
Pupil, gaze, blink, and fixation data were extracted from the recordings through Pupil Labs’ processing techniques applied to the eye and scene videos.
Event triggers were sent at the start and end of each trial, via serial port to the EOG system and via Bluetooth to the eye tracker, enabling temporal alignment across modalities.

\subsection{Data Preprocessing and Feature Extraction}
We used the timestamp of trigger data as the timeline reference and aligned the four EOG channels to it.
Data samples were annotated based on the triggers received.
Following~\cite{wei_electro_2025}, we identified outlier samples as those where the voltage value exceeded three standard deviations from the mean within each channel.
These outliers were replaced using linear interpolation.
Filtering techniques are commonly used to improve signal quality in physiological recordings.
However, filtering operations can introduce edge artifacts, particularly near the boundaries of the signal.
To mitigate these artifacts, we applied a Hanning window with a 0.5-second ramp time.
Afterward, we sequentially removed the 0~Hz frequency component to eliminate the DC offset, applied low-pass filtering to attenuate high-frequency noise, and performed linear detrending to eliminate the trends that typically emerge in biosignals during extended recordings. 
We consistently applied Z-score normalization to the EOG signals recorded during the tasks by scaling the data to zero mean and unit variance, but with different implementations for subject-dependent and subject-independent settings.
In the subject-dependent scenario, we normalized the training data within each subject and used the calculated normalization parameters to normalize the corresponding test data.
However, in the subject-independent scenario, we initially applied Z-score normalization separately to each participant’s dataset to standardize individual variability.
After aggregating the training data across participants, we performed a second normalization and used the resulting parameters to normalize the test data.
Details of the evaluation protocols and dataset splitting are described in the next subsection.
For the EOG features, we used the normalized EOG signals from four channels, along with their differential values.
The differential features were computed by subtracting each sample from its preceding sample, with the first sample initialized to zero to account for the absence of a previous value.

The extracted pupil, gaze, fixation, and blink data are stored separately as nested data structures, with each type organized as a list of dictionaries. 
The dictionaries of each data type contained a varying number of key-value pairs because of different feature sets.
Specifically, the pupil data contained four data streams representing different eyes (eye0, eye1) and measurement methods (2D, 3D). 
We unpacked these data by extracting features into columns within a tabular format.
We also used the event triggers received by the eye tracker to annotate the data samples.
During data alignment, we used the timestamps of the 2D pupil data of eye0 as the reference timeline.
The remaining pupil features, along with the gaze, blink, and fixation features, were synchronized and aligned to this reference. 
Missing pupil and gaze data were filled using the nearest valid values, while missing blink and fixation data were filled with NaN values.
Each eye-tracking data type was accompanied by confidence values reflecting the reliability of its associated measurements.
We identified outliers as data points with confidence values below 0.5.
Contiguous sections of low-confidence data were detected and replaced with NaN values across relevant columns to maintain continuity.
After outlier removal, a Hampel filter with a 100~ms sliding window was applied to replace data points deviating from the median by more than three times the median absolute deviation (MAD) with the median value itself, while ignoring NaN values. 
This was followed by a median filter to further smooth the data.
The eye-tracking data had been normalized by Pupil Labs’ built-in processing.
However, since we used only the segments recorded during the task periods, we applied additional Z-score normalizations to the samples used for evaluation
The normalization procedures for the subject-dependent and subject-independent scenarios followed the same implementation described above for EOG preprocessing.
Confidence values were excluded from subsequent evaluation, as they reflect device-specific information rather than user behavior.
In total, we extracted 74 pupil features, 11 gaze features, 7 fixation features, and 1 blink feature. 
The specific features are listed in~\prettyref{tab:eye_tracking_features}.


\begin{figure*}[ht]
    \centering
    \captionsetup{name={Figure}, justification=justified}
    \includegraphics[width=\textwidth]{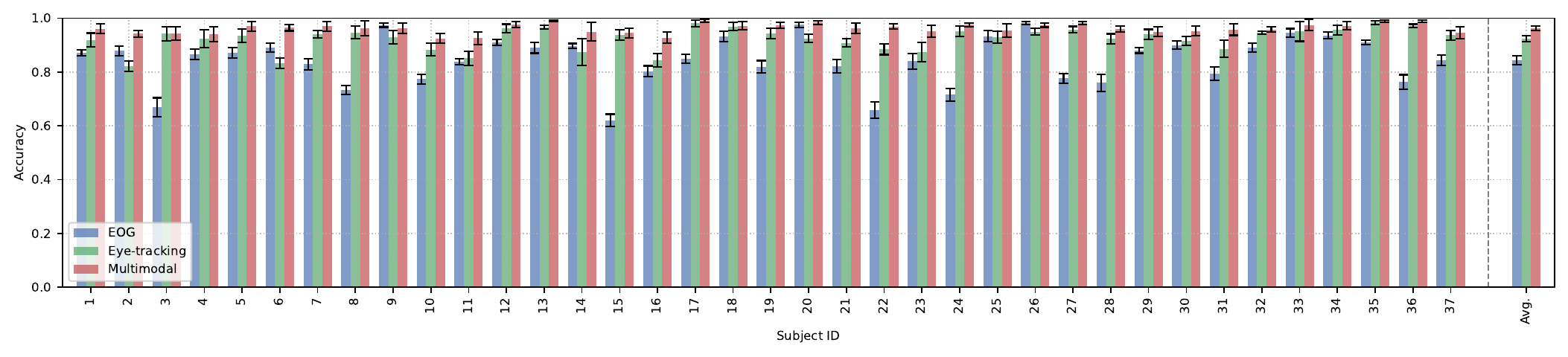}
    \caption{Classification accuracy of EOG, eye-tracking, and multimodal models for each participant in the subject-dependent scenario. Bars represent mean accuracy across eight cross-validation folds per subject, with error bars indicating the standard error of the mean (SEM). On average, the EOG model achieved 84.451\% accuracy, the eye-tracking model 92.432\%, and the multimodal model 96.207\%.}
    \label{fig:acc-dependent}
\end{figure*}

\begin{table*}[t]
\centering
\caption{Macro-averaged F1-scores (\%) of subject-dependent models for all 37 participants, along with the overall average, across models using EOG, eye-tracking, and multimodal configurations.}
\resizebox{\textwidth}{!}{
\begin{tabular}{lcccccccccccccccccccc}
\toprule
Model & P1 & P2 & P3 & P4 & P5 & P6 & P7 & P8 & P9 & P10 & P11 & P12 & P13 & P14 & P15 & P16 & P17 & P18 & P19 \\
\midrule
EOG & $86.8\pm2.9$ & $87.7\pm5.1$ & $66.2\pm9.6$ & $85.9\pm5.7$ & $86.9\pm5.1$ & $88.7\pm4.5$ & $82.5\pm5.7$ & $72.3\pm4.4$ & $97.5\pm2.1$ & $76.9\pm5.0$ & $83.5\pm2.9$ & $90.6\pm3.0$ & $88.9\pm5.4$ & $89.2\pm2.7$ & $61.6\pm5.6$ & $79.5\pm6.2$ & $84.1\pm4.9$ & $92.6\pm5.7$ & $81.0\pm6.6$ \\
ETK & $92.2\pm6.3$ & $82.5\pm6.0$ & $94.1\pm7.4$ & $92.0\pm9.4$ & $93.2\pm6.6$ & $82.7\pm5.9$ & $94.0\pm3.4$ & $94.5\pm6.8$ & $92.5\pm6.9$ & $87.9\pm6.8$ & $84.8\pm7.2$ & $96.0\pm4.6$ & $96.8\pm1.8$ & $87.3\pm13.0$ & $93.8\pm5.5$ & $83.9\pm7.2$ & $98.0\pm3.5$ & $97.0\pm3.6$ & $94.7\pm4.8$ \\
Multi & $96.5\pm3.9$ & $94.2\pm3.3$ & $94.0\pm6.9$ & $93.8\pm7.6$ & $96.9\pm5.0$ & $96.4\pm3.3$ & $96.9\pm5.1$ & $96.1\pm7.9$ & $96.0\pm5.3$ & $92.2\pm5.0$ & $92.2\pm6.7$ & $97.6\pm3.1$ & $99.0\pm0.7$ & $94.9\pm8.9$ & $94.4\pm4.5$ & $92.6\pm5.6$ & $99.1\pm1.6$ & $97.3\pm3.6$ & $97.6\pm2.9$ \\
\midrule
Model & P20 & P21 & P22 & P23 & P24 & P25 & P26 & P27 & P28 & P29 & P30 & P31 & P32 & P33 & P34 & P35 & P36 & P37 & Avg \\
\midrule
EOG & $97.7\pm2.6$ & $81.7\pm6.8$ & $65.2\pm8.1$ & $83.4\pm8.5$ & $71.6\pm5.6$ & $93.0\pm5.7$ & $98.2\pm1.6$ & $77.2\pm4.9$ & $76.0\pm8.8$ & $87.6\pm2.7$ & $89.9\pm4.1$ & $78.5\pm6.4$ & $88.8\pm4.3$ & $94.2\pm4.6$ & $93.3\pm3.3$ & $90.8\pm2.4$ & $75.5\pm7.5$ & $83.6\pm5.0$ & $84.0\pm10.4$ \\
ETK & $92.3\pm4.2$ & $90.6\pm4.5$ & $88.2\pm5.5$ & $86.9\pm9.8$ & $95.2\pm5.2$ & $93.3\pm5.5$ & $95.0\pm3.1$ & $95.8\pm3.1$ & $92.4\pm4.9$ & $93.7\pm5.2$ & $91.5\pm4.5$ & $88.3\pm8.4$ & $94.4\pm1.5$ & $95.1\pm9.7$ & $95.7\pm4.7$ & $98.3\pm2.0$ & $97.1\pm1.9$ & $93.7\pm5.0$ & $92.3\pm7.4$ \\
Multi & $98.4\pm1.8$ & $96.2\pm5.1$ & $96.9\pm2.4$ & $94.8\pm6.3$ & $97.6\pm1.9$ & $95.5\pm5.8$ & $97.5\pm1.9$ & $98.1\pm1.4$ & $96.0\pm3.0$ & $94.6\pm5.0$ & $95.2\pm4.6$ & $95.7\pm5.8$ & $95.6\pm2.7$ & $97.4\pm5.4$ & $97.1\pm4.1$ & $98.9\pm1.1$ & $98.9\pm1.1$ & $94.4\pm6.3$ & $96.1\pm5.1$ \\
\bottomrule
\end{tabular}
}
\label{tab:f1_score_dependent}
\end{table*}

\subsection{Multimodal Refractive Power Estimation}
Before delving into unimodal and multimodal classification, we first describe how data fusion was performed.
As previously described, event triggers were sent simultaneously to both recording systems at the beginning and end of each task.
Therefore, although the internal clocks of the EOG recording devices and the eye tracker were not synchronized, we computed the relative time for each sample by subtracting the timestamp of the trigger received at the start of the task in which the sample was recorded.
This relative timing was then used to align the two modalities.
Due to the sampling rate inconsistency, we downsampled the EOG features from 512~Hz to 120~Hz to align with the eye-tracking features.
This approach ensures that the frequency components of the low-pass filtered EOG signals (with a 50~Hz cutoff) are preserved after downsampling.
Additionally, it avoids upsampling the eye-tracking features, which could amplify measurement noise and introduce artifacts.
Moreover, matching the sampling rates helps maintain consistent sample sizes, thereby enabling a fair comparison between unimodal and multimodal approaches.

We adopted a unidirectional, multi-layer LSTM-based classifier to predict refractive power from eye movement features due to its proven effectiveness in capturing temporal dependencies in time-series classification tasks~\cite{pham_timefrequency_2021}.
We evaluated three models under both subject-dependent and subject-independent scenarios by inputting either a subset of features—8 for the EOG model and 93 for the eye-tracking model—or all 101 features for the multimodal model, using the corresponding input dimensions.
The input eye movement features were first passed through an input projection layer, followed by a ReLU activation, then processed by a four-layer LSTM with 512 hidden units per layer, and finally mapped to class probabilities via a fully connected output layer. 
A dropout rate of 0.5 was applied after the input and LSTM layers to reduce overfitting.
The model output consisted of 13 classes, corresponding to the 13 lenses used in the experiment.
We trained the model for 250 epochs using the Adam optimizer with an initial learning rate of 0.0002. 
A multi-step learning rate scheduler was used, with a milestone at epoch 200 and a decay factor of 0.05. 
The batch size was set to 256.
Cross-entropy loss was used as the objective function.

In the subject-dependent scenario, model training and evaluation were performed within each participant’s dataset.
We employed cross-validation across trials to construct training and testing sets.
As mentioned earlier, each lens condition included 8 fixation trials, 8 pursuit trials, and a 40-second reading task. 
The reading task was divided into 8 equal-length segments to ensure trial-level consistency across conditions.
We created trial-based segments by grouping the following data into a single trial segment: one fixation trial, the inter-trial interval between them, one pursuit trial, and one sub-segment of the reading task.
For each fold of cross-validation, one trial segment was used as the test set, and the remaining seven segments served as the training set.
The training data were further split randomly, with 90\% used for training and 10\% for validation.
In the subject-independent scenario, we adopted a leave-one-subject-out strategy, using each participant’s data as the test set while the data from the remaining 36 participants served as the training set.
This training set was further divided in training and validation sets with 90\% of the data for training and 10\% for validation through random sampling.
We avoided using a leave-one-subject-out strategy for splitting the training and validation sets to prevent overfitting to a specific validation subject.

All evaluations were conducted using PyTorch 2.3.1 on a machine equipped with an NVIDIA GeForce RTX 3080 GPU.
Training subject-dependent models took approximately 36 hours, covering 8 cross-validation folds per participant across three scenarios, resulting in a total of 888 models.
Training subject-independent models took approximately 3 days, with each of the 37 participants serving once as the test subject in a leave-one-subject-out configuration across three scenarios, leading to a total of 111 trained models.

\begin{figure*}[htbp]
  \centering
  \begin{subfigure}[t]{0.32\textwidth}
    \includegraphics[width=\linewidth]{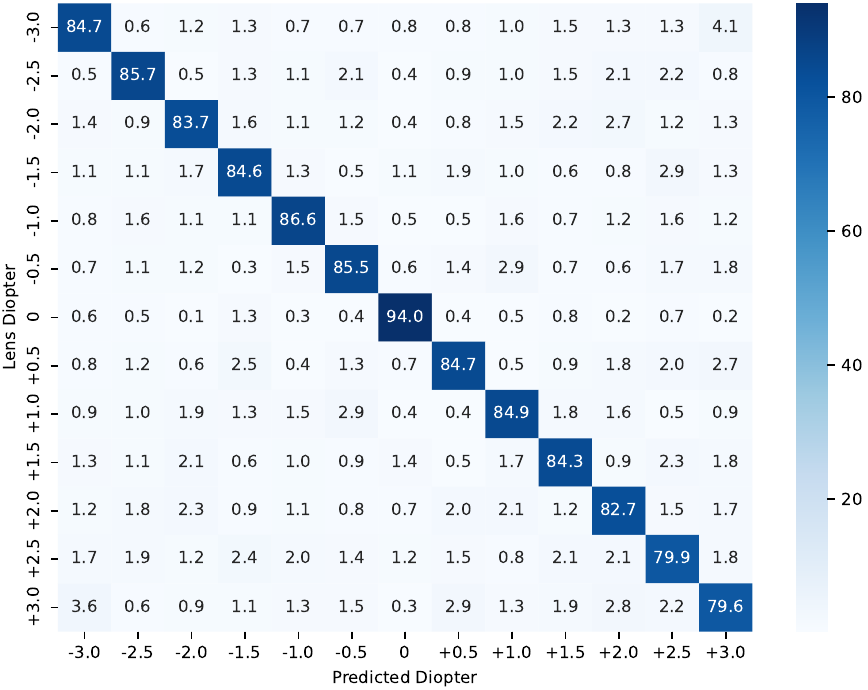}
    \caption{EOG}
    \label{fig:confusion-eog-dependent}
  \end{subfigure}
  \hfill
  \begin{subfigure}[t]{0.32\textwidth}
    \includegraphics[width=\linewidth]{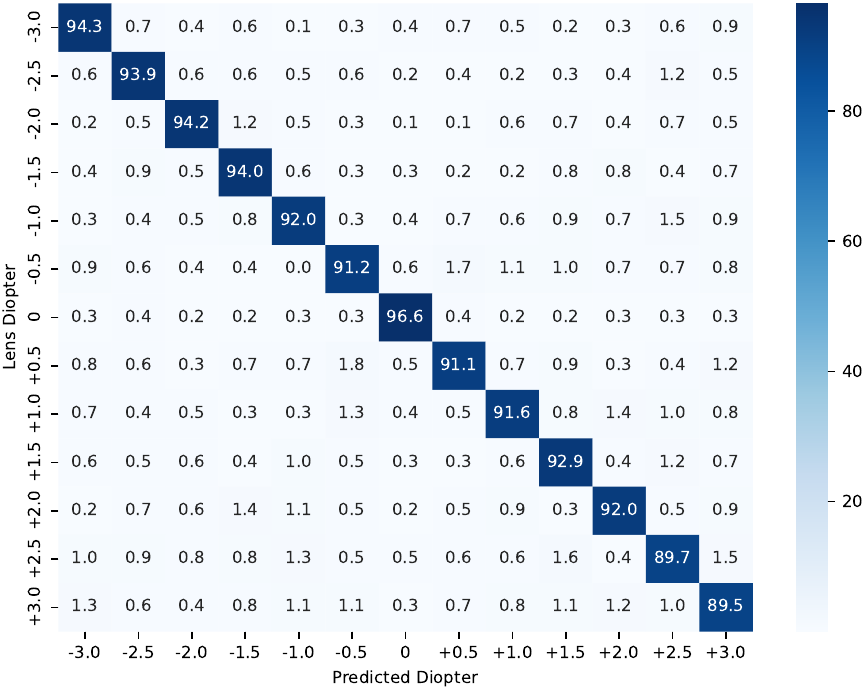}
    \caption{Eye-Tracking}
    \label{fig:confusion-etk-dependent}
  \end{subfigure}
  \hfill
  \begin{subfigure}[t]{0.32\textwidth}
    \includegraphics[width=\linewidth]{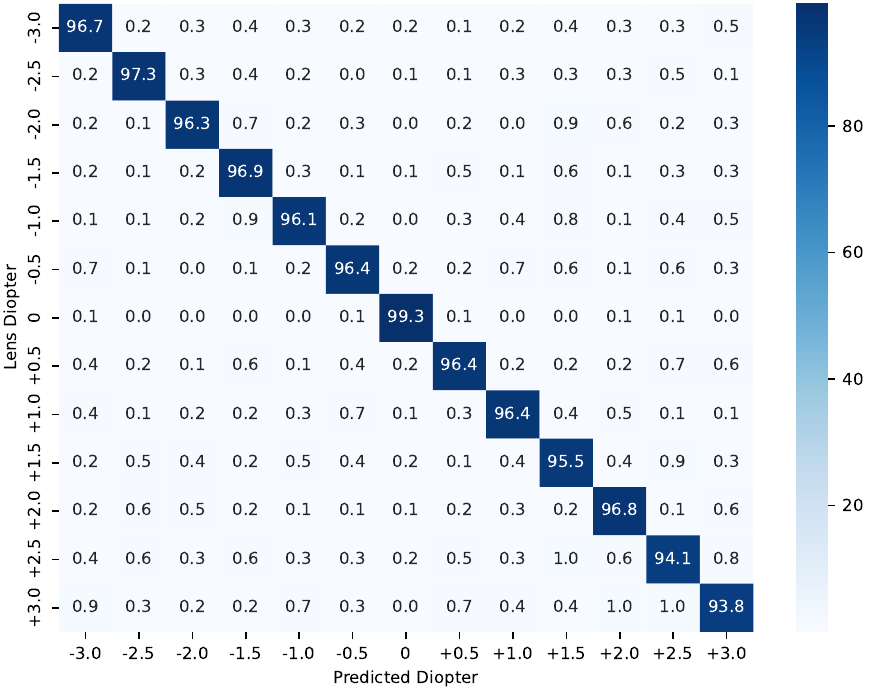}
    \caption{Multimodal}
    \label{fig:confusion-mm-dependent}
  \end{subfigure}
  \caption{Confusion matrices for EOG, eye-tracking, and multimodal models in subject-dependent scenario.}
  \label{fig:confusion-matrices-dependent}
\end{figure*}

\begin{table}[t!]
\centering
\caption{Summary of statistical analyses comparing subject-dependent classification accuracies across models. Bonferroni correction was applied to the p-values obtained from Wilcoxon signed-rank tests.}
\resizebox{\linewidth}{!}{%
\begin{tabular}{l|l|l|l}
\toprule
\textbf{Test Type} & \textbf{Comparison / Model} & \textbf{Statistic} & \textbf{p-value} \\
\midrule
\textbf{Shapiro-Wilk} & EOG            & 0.955 & 0.143 \\
                      & Eye-tracking   & 0.912 & 0.007 \\
                      & Multimodal     & 0.969 & 0.371 \\
\midrule
\textbf{Friedman}      & All models     & 55.622  & $<.001$ \\
\midrule
\textbf{Wilcoxon}     & EOG vs Chance level & 0.0 & $<.001$ \\
                      & Eye-tracking vs Chance level & 0.0 & $<.001$ \\
                      & Multimodal vs Chance level & 0.0 & $<.001$ \\
                      & EOG vs Eye-tracking & 71.0 & $<.001$ \\
                      & Eye-tracking vs Multimodal & 0.0  & $<.001$ \\
                      & EOG vs Multimodal          & 4.0  & $<.001$ \\
\bottomrule
\end{tabular}
}
\label{tab:statistical_results-dependent}
\end{table}

\section{Results}
The classification accuracies of the EOG, eye-tracking, and multimodal models in subject-dependent scenarios are presented in~\prettyref{fig:acc-dependent}.
For each of the 37 participants, accuracy is averaged across eight cross-validation folds, with error bars indicating standard deviation.
The multimodal model, with a mean accuracy of 96.207\%, consistently outperforms the unimodal models across most subjects, achieving higher accuracy and exhibiting greater robustness as indicated by its lower variability.
In addition, the eye-tracking model (mean accuracy = 92.432\%) performed better than the EOG model (mean accuracy = 84.451\%).
On average, the multimodal model achieves the highest classification accuracy, underscoring the advantage of integrating complementary information from both modalities.
~\prettyref{tab:f1_score_dependent} presents macro-averaged F1-scores for all 37 participants.
Each cell shows the average F1-score computed across eight cross-validation folds for the corresponding participant.
The last column summarizes the overall average F1-score for models using EOG, eye-tracking, and multimodal configurations.
A similar trend is observed: the multimodal model again outperforms the unimodal models, achieving the highest average F1-score (96.1\%), compared to eye-tracking (92.3\%) and EOG (84.0\%).
In addition, we presented confusion matrices in~\prettyref{fig:confusion-matrices-dependent}. 
For all three models, most predictions are concentrated along the diagonal, indicating correct classifications.
However, the density and sharpness of the diagonal differ by modality.
A closer examination of the confusion matrices reveals consistent patterns across all three modalities.
While the majority of predictions cluster along the diagonal, indicating generally accurate performance, a notable trend of misclassification emerges between lens conditions with opposite signs but the same diopter magnitude (e.g., -2.0 vs +2.0, -3.0 vs +3.0).
This confusion sometimes extends to their neighboring conditions as well.

To assess whether the performance differences between models were statistically significant, we conducted a series of analyses, including normality testing, a Friedman test, and post-hoc Wilcoxon signed-rank tests.
All statistical results are provided in~\prettyref{tab:statistical_results-dependent}.
The Shapiro–Wilk test showed that the accuracy distributions of the EOG and multimodal models did not significantly deviate from normality ($W = 0.955$, $p = 0.143$ and $W = 0.969$, $p = 0.371$, respectively), while the eye-tracking accuracies violated the normality assumption ($W = 0.912$, $p = 0.007$). 
Based on this, we proceeded with non-parametric analysis.
A Friedman test revealed a significant main effect of modality type on classification accuracy across participants $(\chi^2 = 55.62,\ p < .001)$, indicating that at least one model outperformed the others.
To further investigate, Wilcoxon signed-rank tests were conducted.
All three modality-specific models significantly outperformed the chance level of 7.692\% ($W = 0.0$, $p < .001$). 
Pairwise comparisons also confirmed that the eye-tracking model significantly outperformed the EOG model ($W = 71.0$, $p < .001$), and the multimodal model significantly outperformed both the eye-tracking ($W = 0.0$, $p < .001$) and EOG models ($W = 4.0$, $p < .001$). 
The reported p-values from the Wilcoxon signed-rank tests were adjusted using the Bonferroni correction.

\begin{figure}[t!]
    \centering
    \captionsetup{name={Figure}, justification=justified}
    \includegraphics[width=.7\columnwidth]{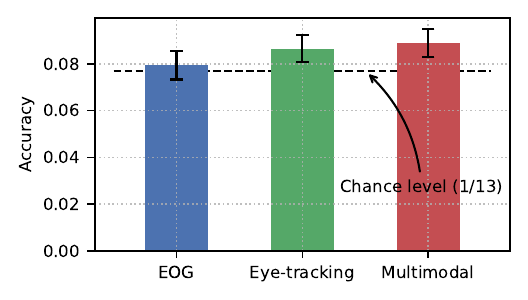}
    \caption{Classification accuracy of EOG, eye-tracking, and multimodal models in the subject-independent scenario. Bars represent mean accuracy across 37 leave-one-subject-out cross-validation folds, with each participant serving once as the test subject. Error bars indicate the standard error of the mean (SEM). On average, the EOG model achieved 7.936\% accuracy, the eye-tracking model 8.640\%, and the multimodal model 8.882\%. The chance level for this 13-class classification task is 7.692\%.}
    \label{fig:acc-independent}
\end{figure}

\begin{table*}[t]
\centering
\caption{Macro-averaged F1-scores (\%) for all subject-independent models, along with the overall average, across models using eog, eye-tracking, and multimodal configurations.}
\resizebox{\textwidth}{!}{
\begin{tabular}{lcccccccccccccccccccc}
\toprule
Model & P1 & P2 & P3 & P4 & P5 & P6 & P7 & P8 & P9 & P10 & P11 & P12 & P13 & P14 & P15 & P16 & P17 & P18 & P19 \\
\midrule
EOG & 10.49 & 6.98 & 5.76 & 5.47 & 2.33 & 6.66 & 10.41 & 5.99 & 2.14 & 5.03 & 11.74 & 7.45 & 5.76 & 1.47 & 9.56 & 7.85 & 6.08 & 2.54 & 5.30 \\
\midrule
ETK & 4.90 & 7.77 & 15.62 & 5.52 & 3.43 & 9.00 & 6.01 & 10.92 & 6.86 & 9.46 & 4.65 & 7.50 & 14.84 & 12.78 & 8.46 & 8.79 & 8.61 & 6.95 & 7.26 \\
\midrule
Multi & 9.87 & 8.07 & 13.84 & 5.23 & 2.31 & 5.97 & 4.79 & 13.78 & 6.94 & 6.81 & 11.18 & 7.31 & 9.03 & 14.57 & 11.61 & 8.00 & 8.31 & 8.55 & 6.69 \\
\midrule
Model & P20 & P21 & P22 & P23 & P24 & P25 & P26 & P27 & P28 & P29 & P30 & P31 & P32 & P33 & P34 & P35 & P36 & P37 & Avg \\
\midrule
EOG & 5.86 & 5.13 & 10.72 & 8.88 & 2.33 & 5.46 & 4.75 & 11.62 & 7.08 & 2.00 & 10.44 & 6.42 & 5.72 & 4.30 & 3.94 & 6.34 & 8.35 & 5.28 & 6.31 \\
ETK & 6.54 & 6.64 & 9.68 & 8.46 & 3.09 & 11.66 & 7.29 & 4.34 & 3.98 & 12.57 & 5.76 & 7.08 & 8.27 & 10.83 & 3.88 & 10.57 & 5.79 & 4.67 & 7.85 \\
Multi & 8.90 & 9.83 & 11.49 & 10.86 & 4.71 & 11.13 & 5.84 & 6.57 & 6.81 & 11.35 & 3.52 & 5.06 & 9.48 & 9.22 & 6.92 & 3.21 & 3.43 & 5.93 & 8.03\\
\bottomrule
\end{tabular}
}
\label{tab:f1_score_independent}
\end{table*}

\begin{figure*}[htbp]
  \centering
  \begin{subfigure}[t]{0.32\textwidth}
    \includegraphics[width=\linewidth]{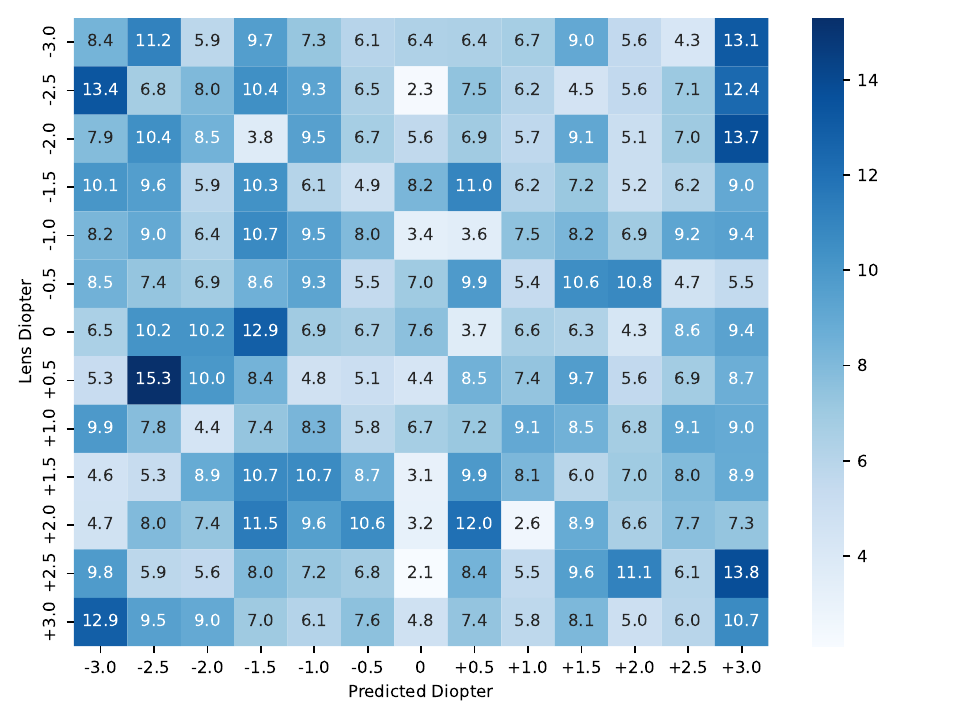}
    \caption{EOG}
    \label{fig:confusion-eog-independent}
  \end{subfigure}
  \hfill
  \begin{subfigure}[t]{0.32\textwidth}
    \includegraphics[width=\linewidth]{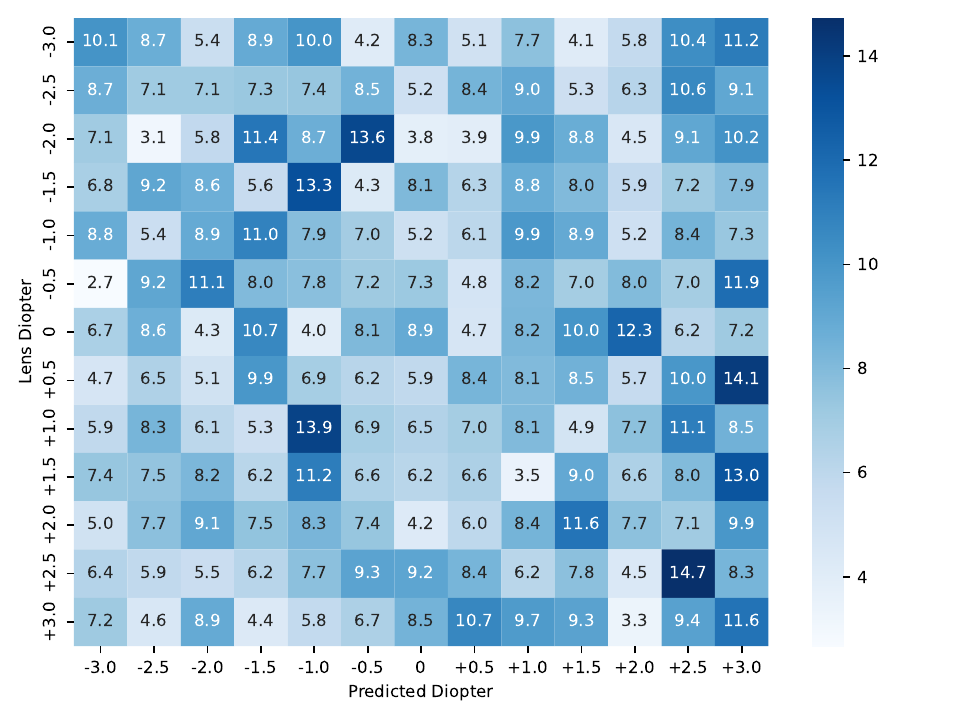}
    \caption{Eye-Tracking}
    \label{fig:confusion-etk-independent}
  \end{subfigure}
  \hfill
  \begin{subfigure}[t]{0.32\textwidth}
    \includegraphics[width=\linewidth]{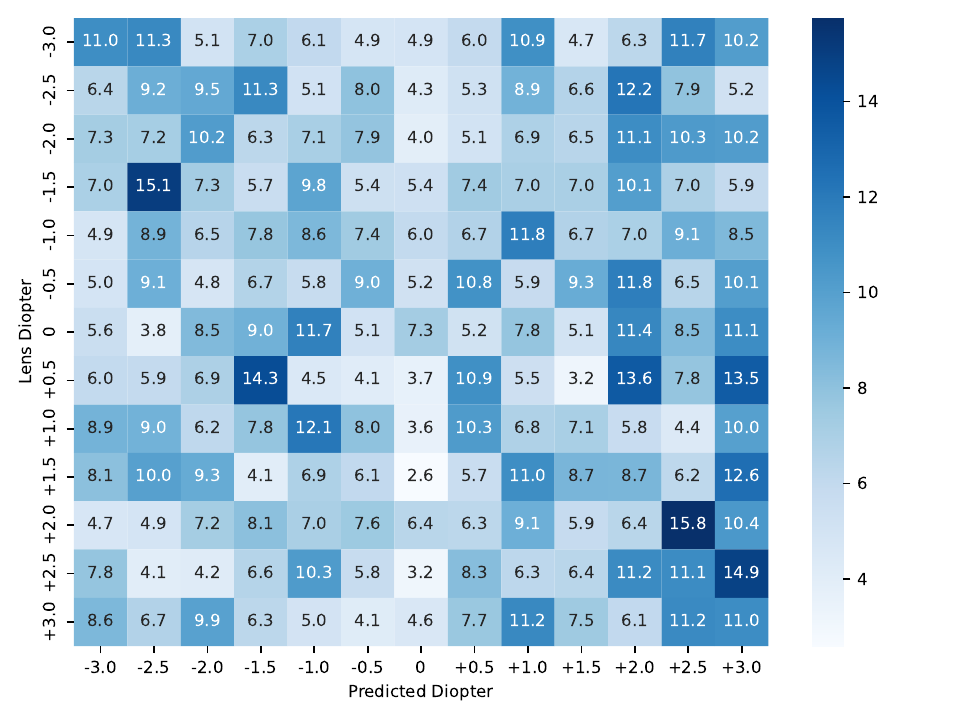}
    \caption{Multimodal}
    \label{fig:confusion-mm-independent}
  \end{subfigure}
  \caption{Confusion matrices for EOG, eye-tracking, and multimodal models in subject-independent scenario.}
  \label{fig:confusion-matrices-independent}
\end{figure*}

\begin{table}[t!]
\centering
\caption{Summary of statistical analyses comparing subject-independent classification accuracies across models.}
\begin{tabular}{l|l|l|l}
\toprule
\textbf{Test Type} & \textbf{Comparison / Model} & \textbf{Statistic} & \textbf{p-value} \\
\midrule
\textbf{Shapiro–Wilk} & EOG           & 0.924 & 0.015 \\
                      & Eye-tracking  & 0.948 & 0.082 \\
                      & Multimodal    & 0.973 & 0.492 \\
\midrule
\textbf{Friedman}      & All models   & 1.459 & 0.482 \\
\bottomrule
\end{tabular}
\label{tab:statistical_results_independent}
\end{table}

\prettyref{fig:acc-independent} illustrates the classification accuracy of the EOG, eye-tracking, and multimodal models in the subject-independent scenario.
The multimodal approach achieved the highest average accuracy (8.882\%), followed by the eye-tracking model (8.640\%) and the EOG model (7.936\%).
Although the overall classification performance in this subject-independent setting is low, the results still highlight the advantage of data fusion through the multimodal approach.
\prettyref{tab:f1_score_independent} reports the corresponding macro F1-scores, which are similar to the accuracy results with the multimodal model achieved the highest average F1-score (8.03\%), followed by eye-tracking (7.85\%) and EOG (6.31\%).
As shown in \prettyref{fig:confusion-matrices-independent}, the confusion matrices show dispersed prediction patterns with no clear diagonal structure or consistent misclassification trends, highlighting the difficulty of generalizing across individuals.

To further analyze the differences in model performance, we conducted statistical tests, as summarized in~\prettyref{tab:statistical_results_independent}.
The Shapiro–Wilk test indicated that the accuracy distributions of the eye-tracking and multimodal models did not significantly deviate from normality ($W = 0.948$, $p = 0.082$ and $W = 0.973$, $p = 0.492$, respectively), while the EOG model showed a statistically significant deviation ($W = 0.924$, $p = 0.015$).
Given these results, we applied a non-parametric Friedman test to compare model performance across participants.
The Friedman test revealed no significant differences among the three models ($\chi^2 = 1.459$, $p = 0.482$).

\begin{table*}[ht]
\centering
\caption{Gaze confidence values (mean ± std) for participants.}
\resizebox{\textwidth}{!}{%
\begin{tabular}{lcccccccccccccccccccc}
\toprule
Subject & P1 & P2 & P3 & P4 & P5 & P6 & P7 & P8 & P9 & P10 & P11 & P12 & P13 & P14 & P15 & P16 & P17 & P18 & P19 \\
\midrule
Confidence & 
$0.85 \pm 0.29$ & $0.87 \pm 0.31$ & $0.91 \pm 0.22$ & $0.99 \pm 0.04$ & $0.92 \pm 0.23$ &
$0.57 \pm 0.44$ & $0.77 \pm 0.29$ & $0.57 \pm 0.46$ & $0.88 \pm 0.27$ & $0.93 \pm 0.20$ &
$0.98 \pm 0.11$ & $0.98 \pm 0.11$ & $0.99 \pm 0.07$ & $0.99 \pm 0.07$ & $0.88 \pm 0.29$ & $0.71 \pm 0.38$ & $0.98 \pm 0.05$ & $0.86 \pm 0.31$ & $0.96 \pm 0.13$ \\
\midrule
Subject & P20 & P21 & P22 & P23 & P24 & P25 & P26 & P27 & P28 & P29 & P30 & P31 & P32 & P33 & P34 & P35 & P36 & P37 & \\
\midrule
Confidence & 
$0.74 \pm 0.38$ & $0.29 \pm 0.39$ & $0.30 \pm 0.35$ & $0.99 \pm 0.04$ & $0.95 \pm 0.15$ &
$0.93 \pm 0.18$ & $0.99 \pm 0.08$ & $1.00 \pm 0.04$ & $0.90 \pm 0.24$ & $0.97 \pm 0.09$ &
$0.85 \pm 0.24$ & $0.90 \pm 0.27$ & $0.99 \pm 0.08$ & $1.00 \pm 0.04$ & $0.94 \pm 0.20$ & $0.95 \pm 0.17$ & $0.88 \pm 0.24$ & $0.89 \pm 0.24$ & \\
\bottomrule
\end{tabular}%
}
\label{tab:confidence}
\end{table*}

\section{Discussion}
In this study, we investigated the estimation of refractive power through eye movement analysis using EOG and video-based eye tracking under varying diopter conditions.
We trained LSTM-based classifiers using unimodal (EOG and eye-tracking) and multimodal inputs and evaluated their performance in both subject-dependent and subject-independent scenarios.
The findings of this study demonstrate both the potential and limitations of using eye movement signals for passive refractive power classification.

In subject-dependent scenario, all three models---EOG, eye-track-ing, and multimodal---demonstrated strong performance.
This suggests that eye movement data contain meaningful information related to refractive power.
The symmetry misclassification pattern observed in the subject-dependent confusion matrices suggests that positive and negative lenses may induce similar eye movement responses, leading the models to confuse these conditions.
Among these methods, the multimodal model achieved the best performance with lower inter-subject variability, suggesting that the integration of EOG and eye-tracking data enhances model predictive power and robustness. 
The eye-tracking model also outperformed the EOG model, which indicates that video-based eye tracking may offer more discriminative features for this task.
However, the unstable quality of the eye-tracking data raises concerns that the discriminative patterns captured under different lens conditions may reflect fluctuations in device stability rather than true differences in eye movement induced by refractive changes. 
To provide a clearer picture, we present the average gaze confidence scores along with their standard deviations in~\prettyref{tab:confidence}.
Participants such as P21 ($0.29 \pm 0.39$) and P22 ($0.30 \pm 0.35$) exhibited notably low and highly variable confidence values. 
In contrast, participants like P5 ($0.99 \pm 0.04$) and P30 ($1.00 \pm 0.04$) showed consistently high confidence, suggesting robust and stable eye-tracking performance. 
These participants with high-quality eye-tracking data also achieved strong classification performance in the subject-dependent setting, supporting the feasibility of using eye-tracking data for refractive power estimation
In contrast, EOG signals were recorded using wet electrodes attached to the skin around the eyes.
This configuration provides a more stable signal quality and is less affected by device-related inconsistencies during the experiment.
Based on these observations, the multimodal approach stands out as the most promising strategy for personalized refractive power estimation.
While eye tracking offers better performance, its signal quality can be unstable.
The reliability and consistency of EOG signals can help mitigate these fluctuations, enhancing overall robustness. 
Given their complementary strengths, integrating both modalities not only boosts performance but also improves robustness in real-world usage.

In the subject-independent scenario, model performance declined, with all models performing only slightly above chance.
Generalizing machine learning models across individuals is a well-recognized challenge when using physiological signals and eye-tracking data~\cite{giakoumis_subject-dependent_2013, huang_discrepancy_2023, wei_unobtrusive_2023, melnyk_gaze_2025}.
Factors such as age, gender, and anatomical differences contribute to this challenge, making it difficult to apply a model trained on one person to another without performance degradation, and even limiting their effectiveness for long-term use on the same individual.
With age, anatomical changes in the eye and alterations in eye movement behavior can influence EOG signal patterns~\cite{grossniklaus_anatomic_2013}.
Older adults tend to exhibit slower saccades, shorter fixation durations, and take longer to respond accurately to visual tasks~\cite{munoz_age-related_1998, ebaid_contribution_2020}.
Across individuals, Miyahira et al. reported significant gender differences in exploratory eye movements, with women exhibiting longer average gaze durations and fewer fixation points, while men showed greater total eye-scanning lengths~\cite{miyahira_gender_2000}.
Variations in eye anatomy, baseline oculometrics, and personal gaze or task strategies can produce inconsistent signal patterns under identical experimental conditions~\cite{unsworth_individual_2019, mehoudar_faces_2014}.
Although we did not observe statistically significant outperformance over the chance level in the subject-independent scenario, the multimodal model still demonstrated relatively higher performance compared to the unimodal models.
This suggests that, despite challenges in generalization, integrating EOG and eye-tracking can still improve classification outcomes, potentially serving as a more robust foundation for future improvements in cross-subject refractive power estimation.
Besides, using personalized models may be a practical direction for developing refractive power estimation tools, especially in health management scenarios where individual calibration, along with ongoing model adaptation based on user-specific data, can significantly enhance reliability and long-term effectiveness.

\section{Limitation and Future Work}
Although this study demonstrates the feasibility of estimating refractive power using eye movement data, several limitations should be noted.
First, the number of extracted features was not balanced between the two modalities. 
The eye-tracking modality contributed substantially more features than EOG, which may have limited the exploration of EOG's discriminative capacity for classifying refractive power and biased the multimodal model toward the more feature-rich modality.
Additionally, the eye-tracking features used in this study were extracted using built-in processing tools provided by the eye-tracker manufacturer, without subsequent feature selection or the extraction of additional features.
This lack of refinement may have limited the model’s ability to fully capture the most informative aspects of the eye-tracking data.
Second, the eye-tracker was not recalibrated before each condition.
As a result, small positional offsets may have accumulated due to session breaks or changes in user posture, introducing variability unrelated to refractive changes.
Combined with the observed instability in eye-tracking quality, these factors may have introduced noise and confounding variables that influenced model performance.

As for future work, several directions can be pursued to address these limitations and improve generalizability.
The current dataset was collected in a controlled lab environment using stationary, non-wearable equipment.
To support real-world applicability, future research should focus on improving the wearability of eye movement data acquisition systems.
Moreover, personalized calibration strategies could be investigated to mitigate inter-subject variability.
For example, transforming subject-specific features into a shared latent representation or normalizing gaze behavior through calibration tasks may enhance model robustness across users.

\section{Safe and Responsible Innovation Statement}
This research involves the analysis of an anonymized dataset comprising electrooculography (EOG) and eye-tracking data collected following the Declaration of Helsinki. Ethical approval was obtained before data collection.
The eye tracker was calibrated once at the start of the session and not recalibrated after breaks. Although some participants remained on the chinrest, slight offsets may have been introduced due to variations in head placement.
The eye-tracking results are based on data collected using a specific device, and generalizability to other systems may vary.
Given that the physical characteristics of the human eye can change over time, the models developed in this study may not remain valid for long-term use.

\bibliographystyle{ACM-Reference-Format}
\bibliography{reference}

\end{document}